\documentclass[twocolumn, prd, aps, nofootinbib, superscriptaddress]{revtex4-1}
\usepackage[utf8]{inputenc}
\usepackage[english]{babel}
\usepackage{amsmath}
\usepackage{mathrsfs}
\usepackage{amssymb}
\usepackage{graphicx}
\usepackage{physics}
\usepackage{hyperref}
\usepackage{xcolor}
\usepackage{comment}

\def\aap{\rm AAP}

\def\apj{\rm ApJ}
\def\apjl{\rm ApJL}

\def\jcap{J. Cosm. Astro. Phys.}
\def\mnras{\rm MNRAS}
\def\nat{\rm Nature}

\def\physrep{\rm Phys. Rep.}

\begin{document}

\title{Investigating the I-Love-Q and $w$-mode Universal Relations Using Piecewise Polytropes}

\author{Ernesto Benitez}
\affiliation{Department of Astronomy, University of Maryland, College Park, Maryland 20742, USA}
\affiliation{Department of Physics, University of Maryland, College Park, Maryland 20742, USA}
\author{Joseph Weller}
\affiliation{Department of Astronomy, University of Maryland, College Park, Maryland 20742, USA}
\affiliation{Department of Physics, University of Maryland, College Park, Maryland 20742, USA}
\author{Victor Guedes}
\affiliation{Center for Natural and Human Sciences, UFABC, Santo Andr{\'e}-SP, 09210-170, Brazil}
\author{Cecilia Chirenti}
\affiliation{Department of Astronomy, University of Maryland, College Park, Maryland 20742, USA}
\affiliation{Astroparticle Physics Laboratory NASA/GSFC, Greenbelt, Maryland 20771, USA}
\affiliation{Center for Research and Exploration in Space Science and Technology, NASA/GSFC, Greenbelt, Maryland 20771, USA}
\affiliation{Center for Mathematics, Computation and Cognition, UFABC, Santo Andr{\'e} - SP, 09210-580, Brazil}
\author{M. Coleman Miller}
\affiliation{Department of Astronomy, University of Maryland, College Park, Maryland 20742, USA}

\date{\today}

\begin{abstract}
Neutron stars are expected to have a tight relation between their moment of inertia ($I$), tidal deformability  ($\lambda$, which is related to the Love number), and rotational mass quadrupole moment ($Q$) that is nearly independent of the unknown equation of state (EoS) of cold  dense matter.  These and similar relations are often called ``universal", and they have been used for various applications including 
analysis of gravitational wave data.  We extend these studies using piecewise polytropic representations of dense matter, including for so-called twin stars that have a second branch of stability at high central densities.  The second-branch relations are less tight, by a factor of $\sim 3$, than the relations found in the first stable branch.  We find that the relations on both branches become tighter when we increase the lower limit to the maximum mass for the EoS under consideration.  We also propose new empirical relations between $I$, $\lambda$, $Q$, and the complex frequency $\omega=\omega_R+i\omega_I$ of the fundamental axial $w$-mode, and find that they are comparably tight to the I-Love-Q correlations.  

\end{abstract}

%\keywords{}

\maketitle

\section{Introduction}

The properties of the matter in the cores of neutron stars are not well known, because the relevant densities and neutron-proton asymmetries cannot be explored in laboratories and because observations of neutron stars are not yet precise enough to be definitive (see \cite{2001ApJ...550..426L,2013arXiv1312.0029M,2016EPJA...52...63M,2020ApJ...888...12M}; but see \cite{2017PhRvL.119p1101A,2018PhRvL.121i1102D,2019ApJ...887L..21R,2019ApJ...887L..24M} for recent measurements from gravitational waves and X-ray observations).  Nonetheless, numerous studies using tabulated equations of state (EoS: the pressure as a function of the energy density) have shown that there are macroscopic properties of neutron stars that are tightly correlated with each other in a way that is insensitive to the detailed physics of the cores.  For example, nearly independent of the unknown EoS of matter beyond nuclear saturation density, knowledge for slowly rotating neutron stars of the moment of inertia ($I$), tidal Love number (Love, or $\lambda$), or rotational mass quadrupole moment ($Q$) implies knowledge of the other two to within $\sim 1-2$\% (e.g., \cite{2013Sci...341..365Y,2013PhRvD..88b3009Y,2013PhRvD..88b3007M,2014PhRvD..90f3010Y,2014PhRvL.112l1101P,2017PhR...681....1Y,2019PhRvD..99h3016C}; note however that the relations become much less tight for rapidly rotating neutron stars \cite{2014ApJ...781L...6D} (although a good correlation can be reestablished using a suitable change of variables \cite{2014PhRvL.112t1102C}) or stars with strong internal magnetic fields \cite{2014MNRAS.438L..71H}).  
%Although none of $I$, $\lambda$, or $Q$ have been measured for any neutron star (upper limits to $\lambda$ were obtained for the double neutron star coalescences GW170817 \cite{2017PhRvL.119p1101A} and GW190425 \cite{2020ApJ...892L...3A}), the so-called I-Love-Q relation has been used to obtain improved precision in tidal deformability constraints from the double neutron star merger event GW170817 \cite{2018PhRvD..97h4038P} and may in the future be used for new tests of general relativity.

None of $I$, $\lambda$, or $Q$ have been measured for any neutron star. However,  the so-called I-Love-Q relation has been used to obtain improved precision in tidal deformability constraints from the double neutron star merger events (GW170817 \cite{2018PhRvD..97h4038P} and GW190425 \cite{2020ApJ...892L...3A}), and may in the future be used for new tests of general relativity.

So far, tests of the I-Love-Q relation have largely been confined to a limited set of tabulated EoS, and most studies have focused entirely on $I$, $\lambda$, and $Q$ rather than other potentially correlated quantities (see \cite{2018CQGra..35a5005S,2019PhRvD..99j4005M} for exceptions to these rules).  Moreover, a relatively unexplored region of parameter space is that of second branches of stability, i.e., EoS that produce stable stars up to a certain central density, then unstable stars up to another threshold density, then stable stars again for a set of yet higher densities.  These have been studied under the name of ``twin stars" (and are also considered to be a third family of degeneracy-supported objects, where white dwarfs form the first family), and although nature might not select such EoS they are interesting because, for example, if such stars exist it is possible that two stars could have the same gravitational mass but significantly different radii (e.g., \cite{2000A&A...353L...9G,2015PhRvD..92h3002A,2015A&A...577A..40B,2017PhRvC..96d5809A,2017PhRvC..96f5807R,2018ApJ...860..139B,2018JCAP...12..031R,2019PhRvD..99j3009M,2019arXiv190602522B,2019PhRvD..99h3014H,2019PhRvC.100b5802M,2019JPhG...46g3002O}).

Here we: (1)~parameterize the high-density EoS using many realizations of a five-segment piecewise polytrope, (2)~explore correlations of the real and imaginary components of the frequency of $w$-modes (a type of spacetime mode; see Section~\ref{sec:w-mode}) with $I$, $\lambda$, and $Q$, and (3)~generate and analyze stars in second stable branches.  We find that the overall I-Love-Q correlation is strong, although it has greater dispersion in the second branch.  We also find that the $w$-mode frequencies have correlations with $I$, $\lambda$, and $Q$ that are comparably tight to the correlations that the three have with each other.

\section{Methods}
\label{sec:methods}

Our goal is to compute $I$, $\lambda$, $Q$, and the $w$-mode frequencies for a large number of simulated neutron stars.  The first step in the computation is to choose an EoS and a central density.  To do this we use the following procedure, which is common in the field:

\begin{figure}[ht]
        \includegraphics[width=\linewidth]{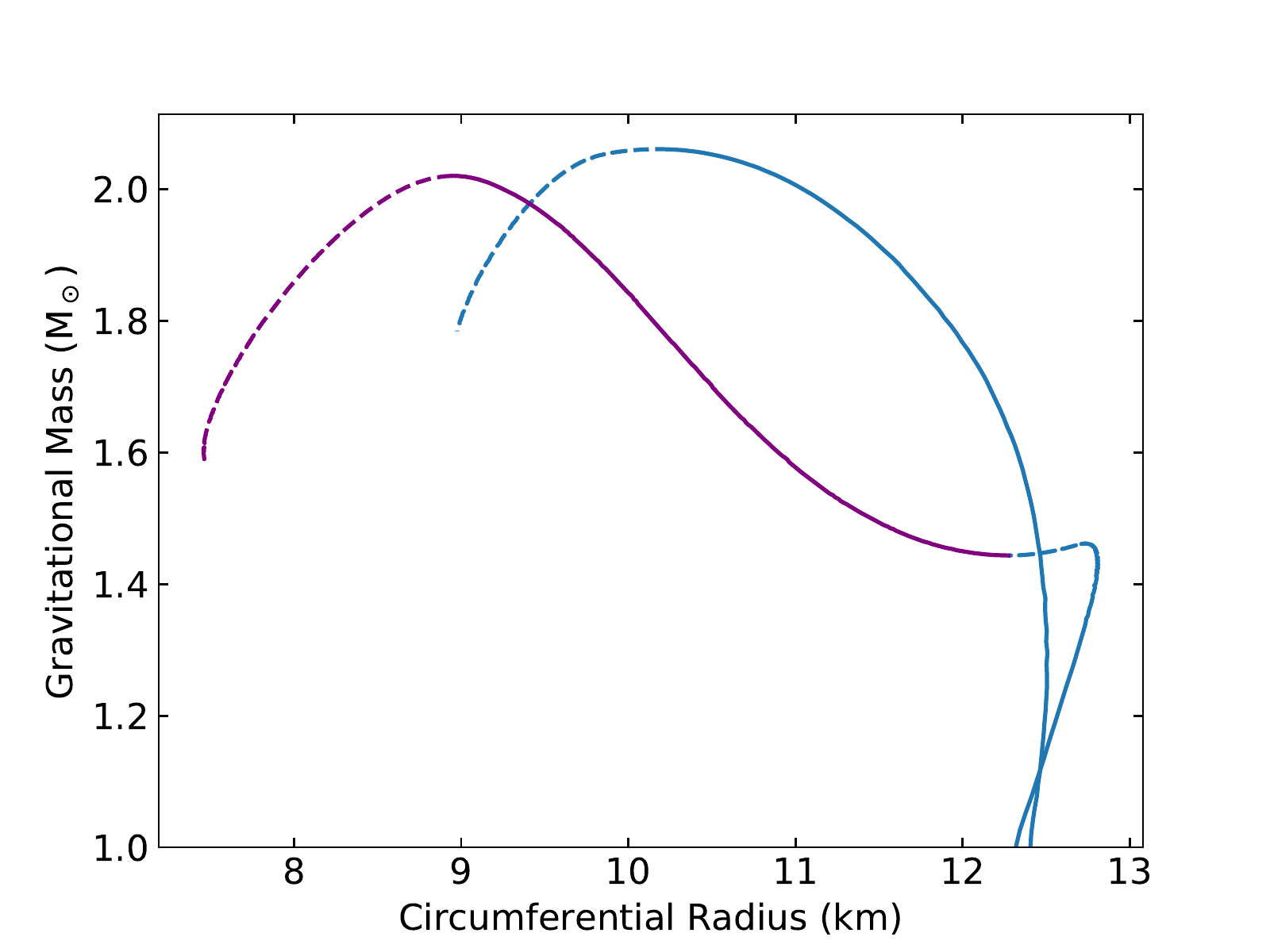}
    \caption{Example mass-radius relations for the equations of state we consider.  The solid blue parts of the curves show the first branch of stability, and the solid magenta part of one of the curves shows the second branch of stability.  The dotted portions of the curves show regions that are unstable. The central density of the star increases along each curve from the bottom right to the top left.}
    \label{fig:MR}
\end{figure}

\begin{enumerate}

\item We assume that we know the EoS up to half of nuclear saturation density, i.e., to a baryonic rest mass density $\rho=\rho_s/2\approx 1.34\times 10^{14}$~g~cm$^{-3}$.  This is justified because at such low densities laboratory experiments can provide guidance (although at a much lower neutron to proton ratio than in a neutron star).  Commonly the EoS used at low density is SLy4 \cite{2001A&A...380..151D}, but here we use the more updated EoS QHC19 \cite{2019ApJ...885...42B}.  Because we assume a standard hadronic EoS at low densities, this means that our approach cannot model self-bound strange stars \cite{2010PhRvD..82b4016P}.

\item Above half nuclear density, we parameterize the EoS.  There are many possible approaches (see \cite{2009PhRvD..79l4032R,2010PhRvD..82j3011L,2014ApJ...789..127K,2016ApJ...831...44R,2018PhRvL.120q2703A,2018PhRvL.120z1103M,2019PhRvD..99h4049L,2020NatPh..16..907A,2020arXiv200803342O} for examples and applications).  Here we choose a piecewise polytropic parametrization with five density segments, following the recommendation of \cite{2016ApJ...831...44R}, although we do not expect our results to be sensitive to the choice of parameterization.  The transition densities are $\rho_s$, $2\rho_s$, $4\rho_s$, and $8\rho_s$, and in segment $i$ the pressure $P$ is given by $P(\rho)=K_i\rho^{\Gamma_i}$.  The polytropic indices $\Gamma_i$ are selected uniformly from 0 to 5, and the coefficients $K_i$ are chosen to enforce continuity of the pressure at the transition densities.  If at any density the implied adiabatic sound speed $c_s=(dP/d\epsilon)^{1/2}>c$ (where $c$ is the speed of light and $\epsilon$ is the total energy density) then we set $c_s=c$ for that density.

\item For a given EoS, we need to choose a central density, which serves as our boundary condition for integration of the equations of stellar structure.  For our first-branch set of stars we choose this central density uniformly between what would produce a $1.0~M_\odot$ star, and the highest density in the first stable branch (or the single stable branch if there is only one).  For our second-branch set of stars, we choose a central density uniformly between the lowest and highest densities of the second stable branch.  Example mass-radius curves for both types are shown in Figure~\ref{fig:MR}.  All of our stars have a maximum mass of at least $1~M_\odot$.

\item We assume that our stars rotate slowly enough that we can use the Tolman-Oppenheimer-Volkoff \cite{1939PhRv...55..364T,1939PhRv...55..374O} equation for nonrotating stars.

\end{enumerate}

We have 497,220 stars in our first-branch set; this compares with 27,440 in \cite{2019PhRvD..99j4005M} (all of which are first-branch stars).  We also have 9,484 stars in our second-branch set. Although our methodology does not explicitly consider changes in composition, second branches arise in EoS that have combinations of polytropic indices that mimic phase transitions, that is, those having $\Gamma_i \approx 0$ in some density segment $i$ (see \cite{2017PhRvL.119p1104A,2020arXiv200903769R} for the implications of multiple phase transitions).

The second step in the computation is the calculation of $I$, $\lambda$, $Q$, and the $w$-mode frequencies.  To calculate $I$ and $Q$ we follow the treatment of \cite{1967ApJ...150.1005H} including the correction noted in \cite{1968ApJ...153..807H} (see their Equation~(26) and the associated footnote 5).  To compute $\lambda\equiv \frac{2}{3}G^{-1}R^5k_2$ (where $G$ is Newton's constant, $R$ is the circumferential radius, and $k_2$ is the $l=2$ electric tidal Love number, or apsidal constant) we follow \cite{2008ApJ...677.1216H}, as amended in the erratum \cite{2009ApJ...697..964H}.  The relevant equations are summarized compactly in \cite{2013PhRvD..88b3009Y}.  Solution of the equations for $I$ and $Q$ requires the assumption of a small but nonzero angular velocity at the center of a uniformly rotating star; in practice, if the assumed angular velocity is too small then significant numerical errors are possible.  We therefore use a central angular velocity of 10~Hz and have confirmed that moderately different choices do not lead to significantly different values for $I$ and $Q$.

The $w$-modes are a class of odd-parity (also called axial) nonradial perturbations that are spacetime modes of neutron stars \cite{1991RSPSA.434..449C,1992MNRAS.255..119K,1994MNRAS.268.1015K}. Because these modes decay and are therefore quasinormal modes, they have a real and an imaginary component to their frequencies: $\omega=\omega_R+i\omega_I$, where $\omega_I$ is the inverse of the decay time (see \cite{1999LRR.....2....2K,1999CQGra..16R.159N} for reviews on black hole and neutron star quasinormal modes).  Early papers investigated the relation of the $w$-mode complex frequency with the stellar compactness $GM/(Rc^2)$, where $M$ is the gravitational mass of the star \cite{1998MNRAS.299.1059A,1999MNRAS.310..797B,2005MNRAS.357.1029T}.  Later work explored the effect of rescaling the $w$-mode frequencies with the square root of the central pressure \cite{2013PhRvD..87j4042B}, and correlations of the rescaled frequencies with $\lambda$ \cite{2019PhRvD..99j4005M}.  Here, to compute the $w$-mode frequencies, we follow the numerical scheme of \cite{2019PhRvD..99j4005M}.  We find it advantageous to employ a shooting technique, by which we match, at the surface of the star, an outward-integrated solution (imposing regularity at the center of the star) with an inward-integrated solution (imposing an outgoing wave solution at infinity, using the exterior complex scaling technique first proposed by \cite{1995MNRAS.274.1039A}).

Once we have computed $I$, $\lambda$, $Q$, and the $w$-mode frequencies for each of our neutron stars, we follow \cite{2013PhRvD..88b3009Y} in fitting a fourth-order polynomial to our values:
\begin{equation}
Y_i=c_{0i}+c_{1i}X+c_{2i}X^2+c_{3i}X^3+c_{4i}X^4\; ,
\label{eq:4orderfit}
\end{equation}
where $c_{0i}, c_{1i}$ etc. are our fitting coefficients, we take $\log_{10}{\bar Q}$ as our independent variable $X$, and $Y_i$ is $\log_{10}{\bar I}$, $\log_{10}{\bar\lambda}$, or the real or imaginary part of the $w$-mode frequency.  Here ${\bar Q}\equiv-Q/(\chi^2M^3)$, ${\bar I}\equiv I/M^3$, and ${\bar\lambda}\equiv \lambda/M^5$ are respectively the dimensionless rotational mass quadrupole, dimensionless moment of inertia, and dimensionless tidal deformability, and $\chi\equiv J/M^2$ (for stellar angular momentum $J$) is the dimensionless angular momentum, in units where $G=c\equiv 1$.  In the next section we perform these fits using different portions of our data sets and plot residuals to the fits.

%\section{Analysis}

\begin{figure*}[ht]
        \includegraphics[width=0.49\textwidth]{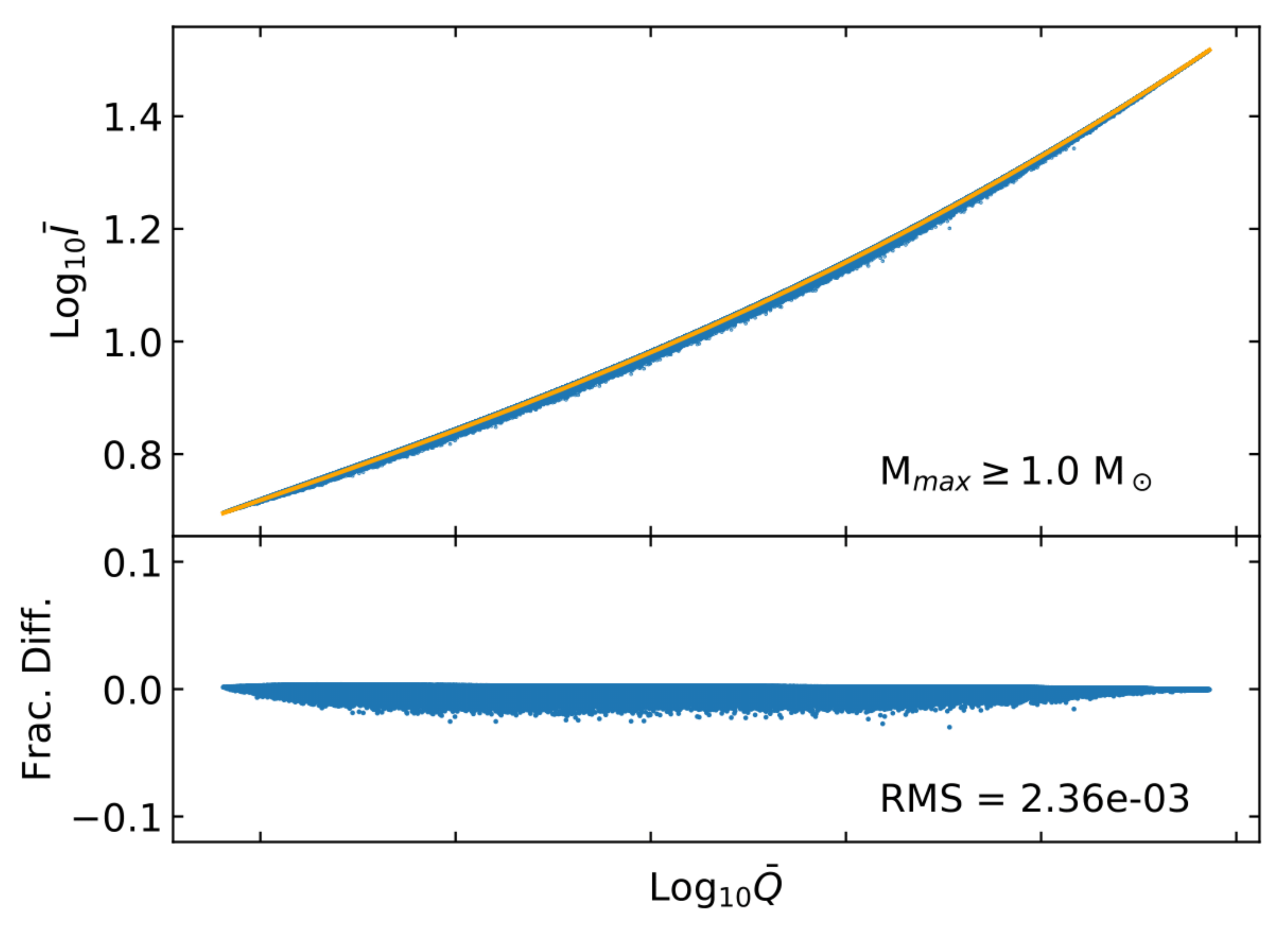}
        \includegraphics[width=0.49\textwidth]{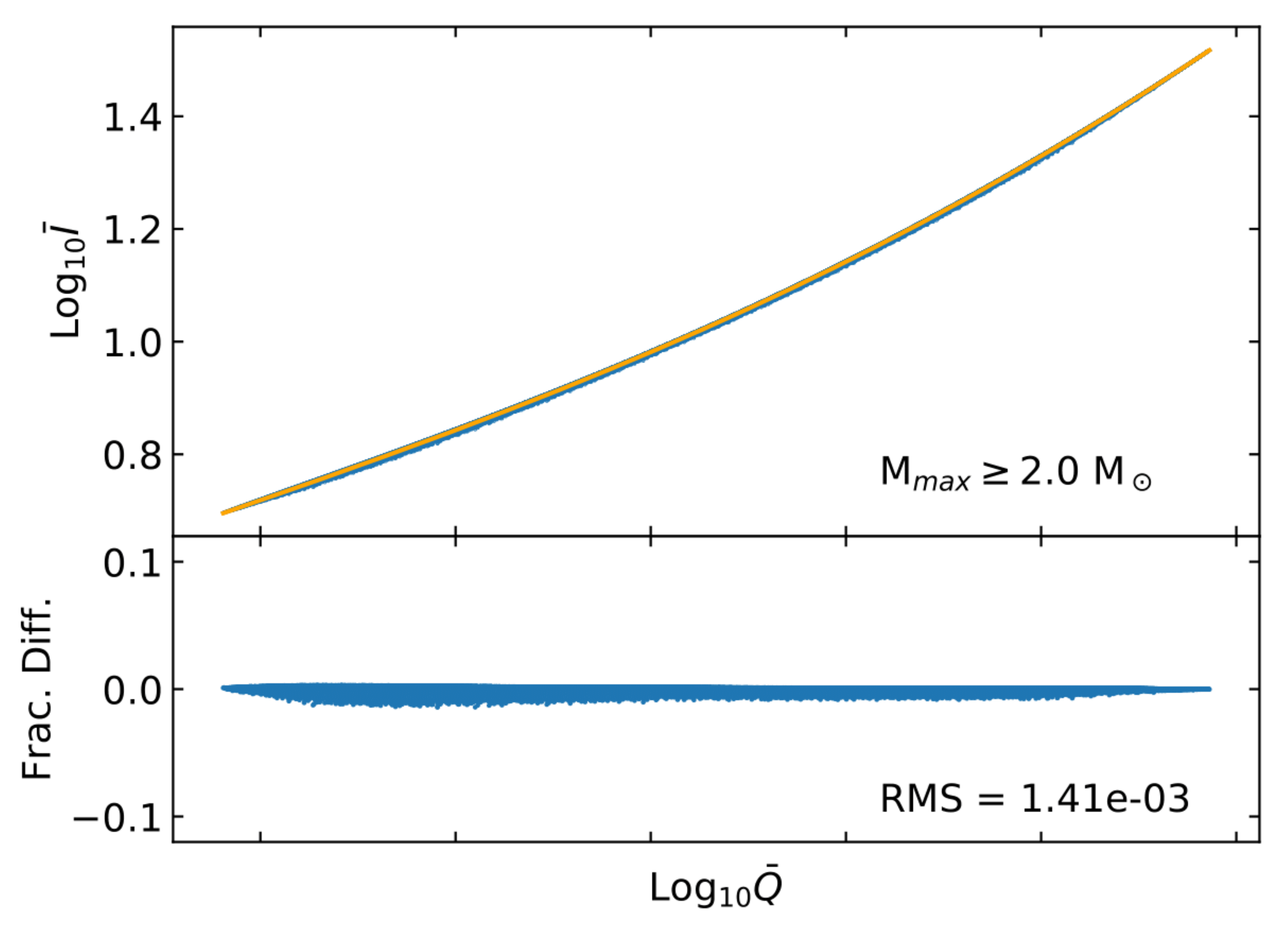}
        \includegraphics[width=0.49\textwidth]{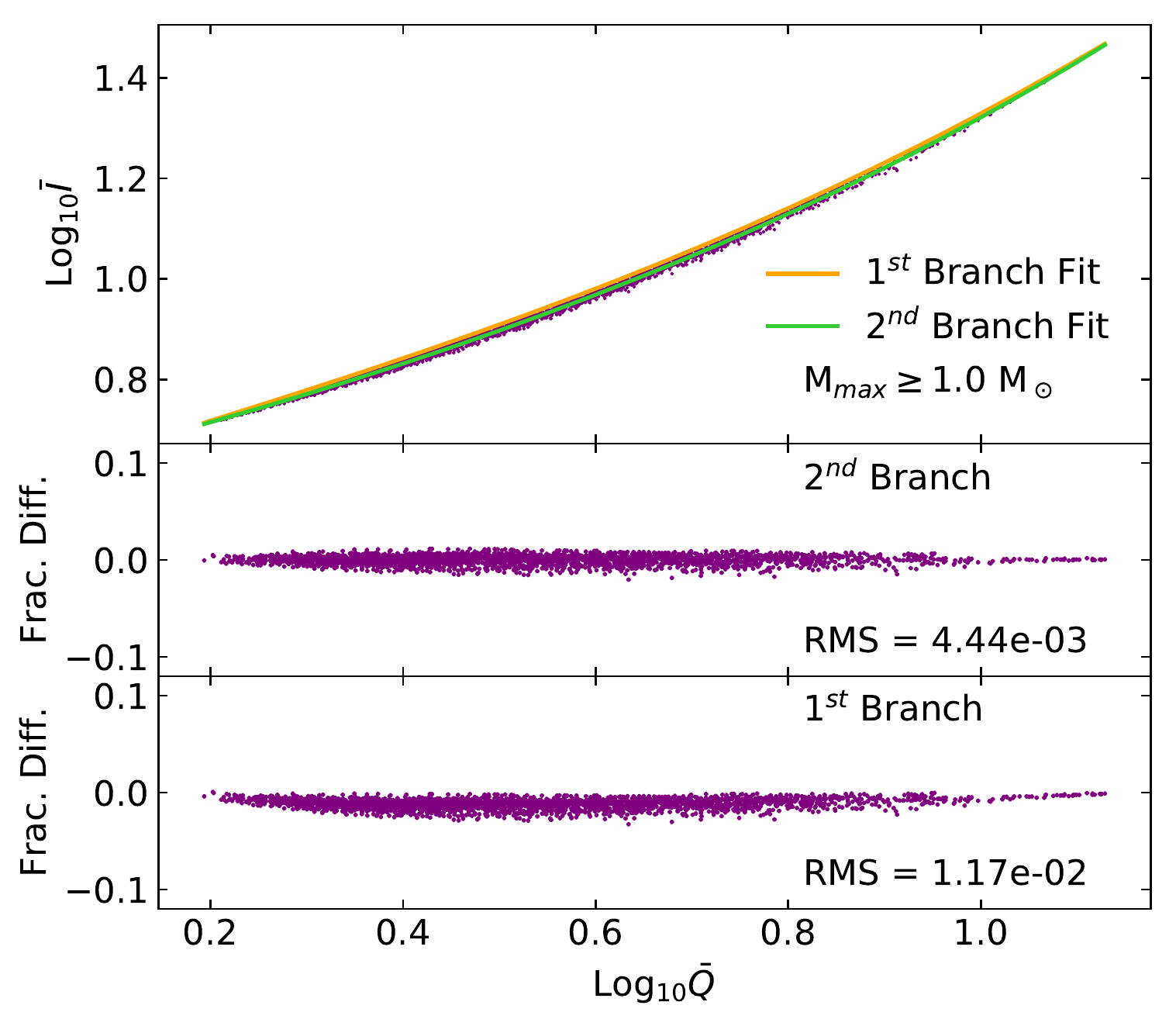}
        \includegraphics[width=0.49\textwidth]{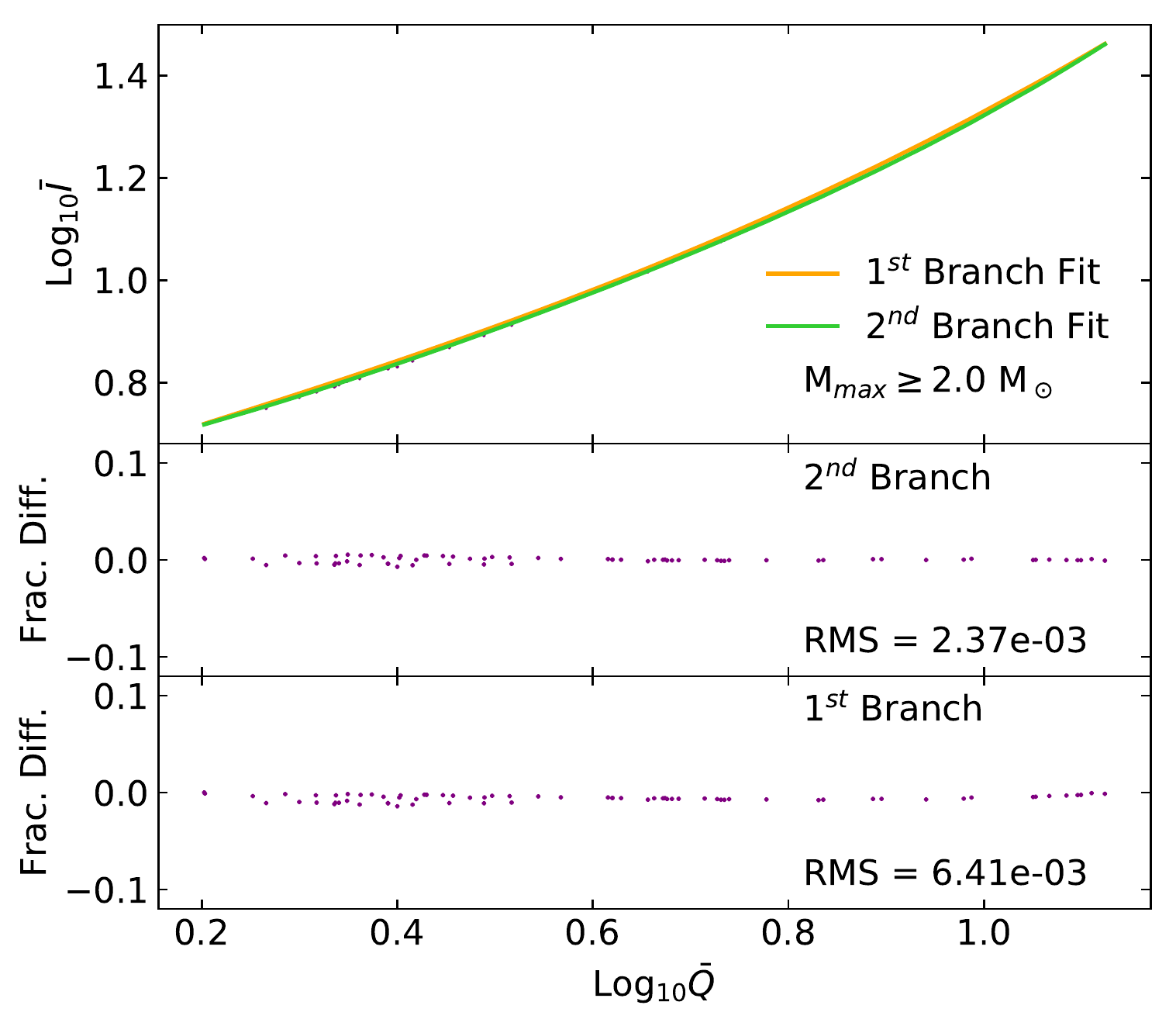}
    \caption{The $\log_{10}{\bar Q}-\log_{10}{\bar I}$ relation and its residuals.  Top left: our full first-branch set of neutron stars.  Top right: only the first-branch stars constructed using EoS with maximum gravitational mass $\geq 2~M_\odot$.  In both of the top figures, the bottom panel shows the fractional difference from the best fourth-order fit (see Equation~\ref{eq:4orderfit}), as well as the overall root-mean-square (rms) deviation.  Bottom left: only the stars constructed from a second high-density stable branch.  Bottom right: only the stars constructed from a second high-density stable branch using an EoS with a maximum gravitational mass $\geq 2~M_\odot$.  For the bottom figures, the middle panel shows the fractional differences and rms relative to our fourth-order fit to the second-branch data, and the bottom panel shows the fractional differences and rms relative to our fourth-order fit to the first-branch data.  We see that increasing the minimum value of the maximum mass tightens the relations for the second-branch stars as well as for the first-branch stars.  We also see that the second-branch stars follow a slightly different relation than the first-branch stars.}
    \label{fig:QIPlots}
\end{figure*}

\section{Results}
\label{sec:results}

\subsection{I-Love-Q}

\begin{figure*}[ht]
        \includegraphics[width=0.49\textwidth]{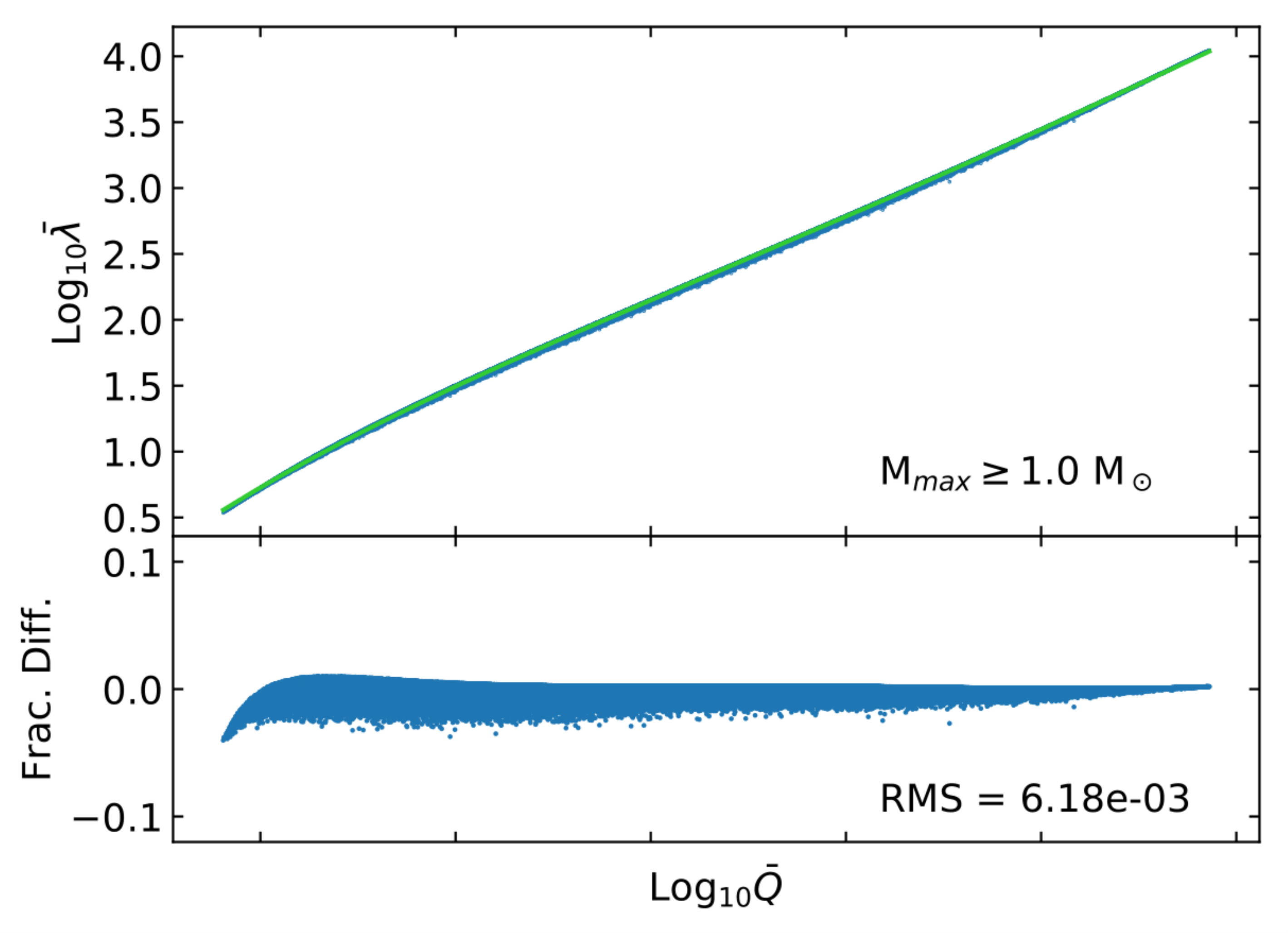}
        \includegraphics[width=0.49\textwidth]{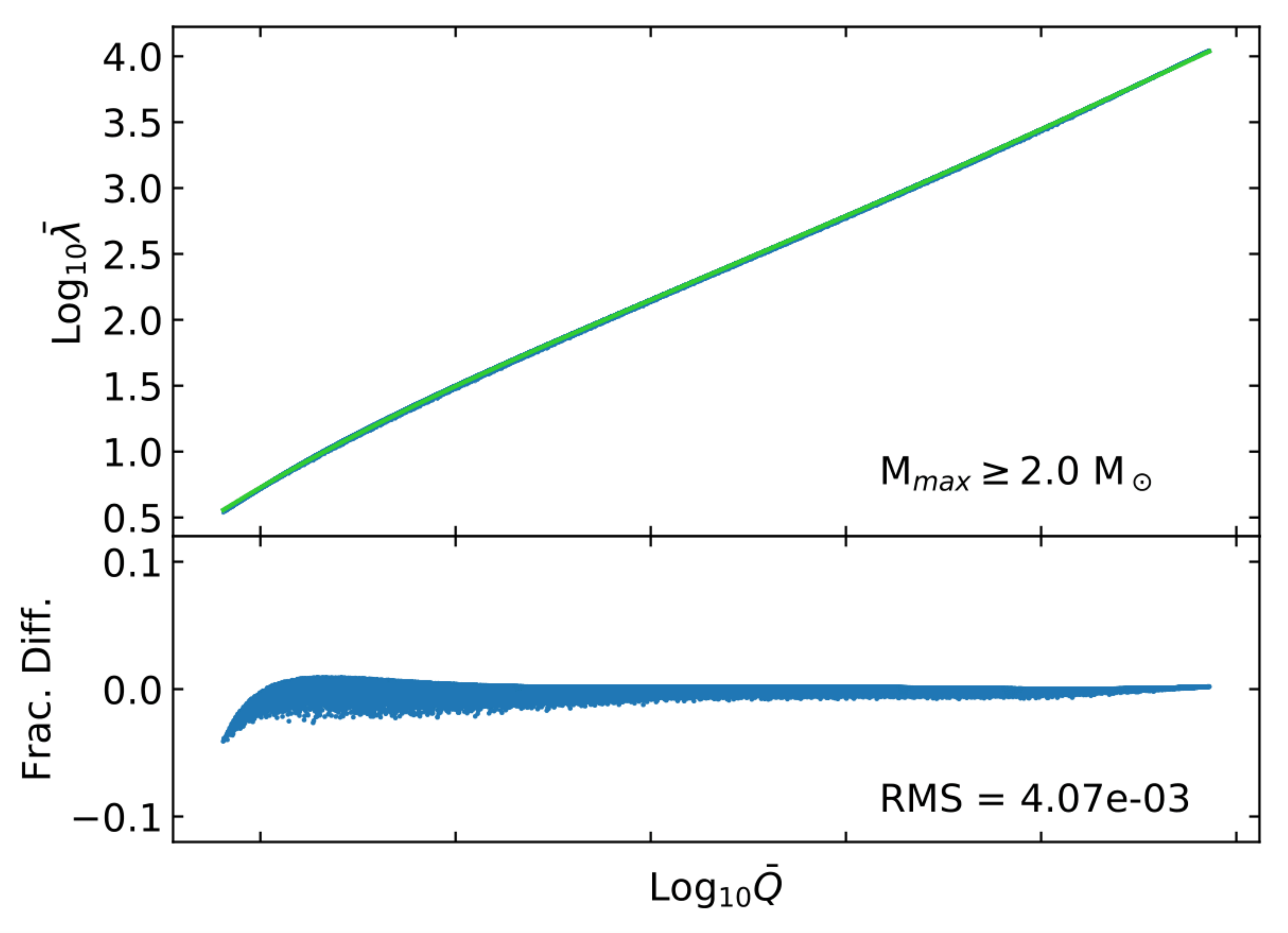}
        \includegraphics[width=0.49\textwidth]{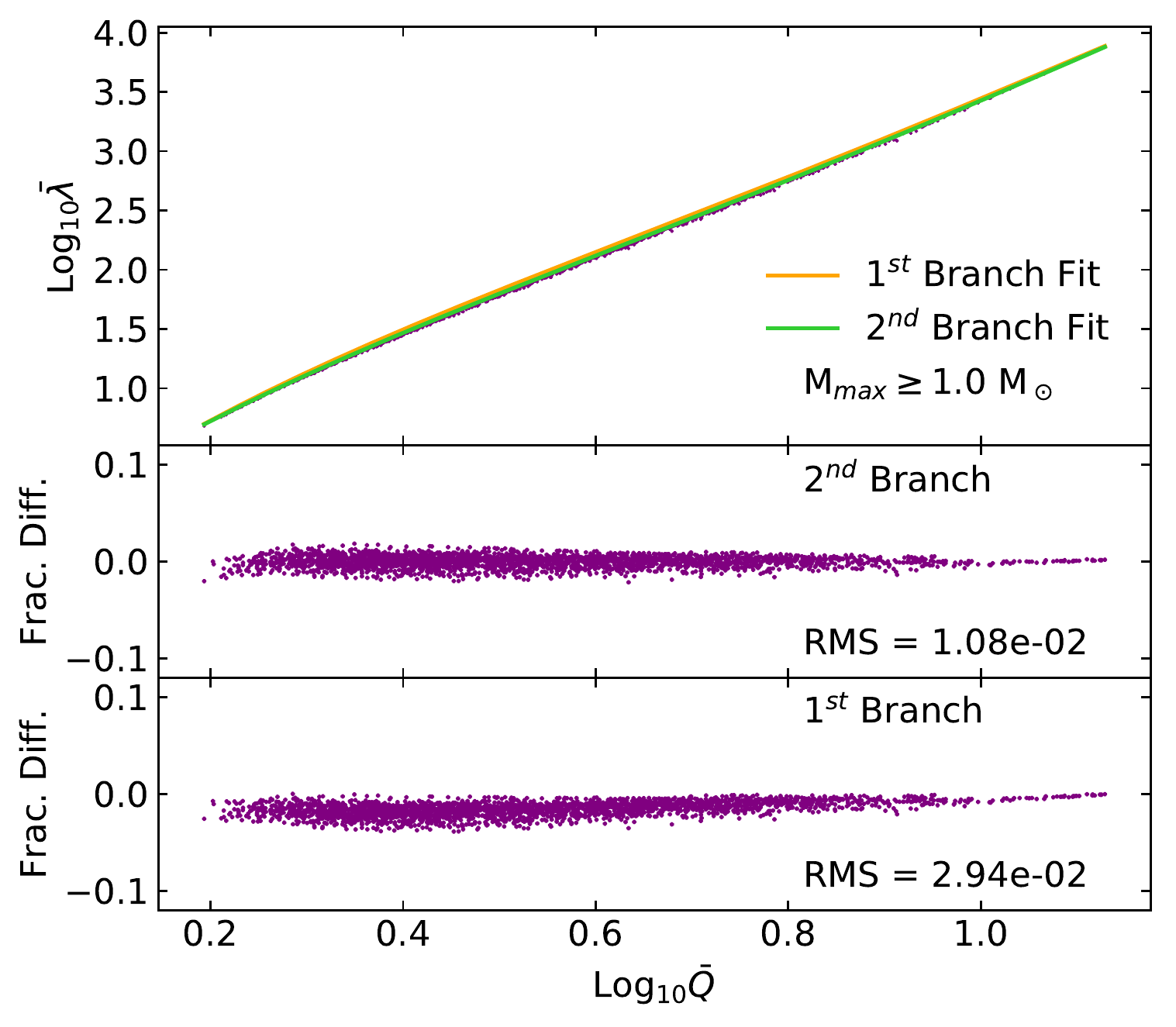}
        \includegraphics[width=0.49\textwidth]{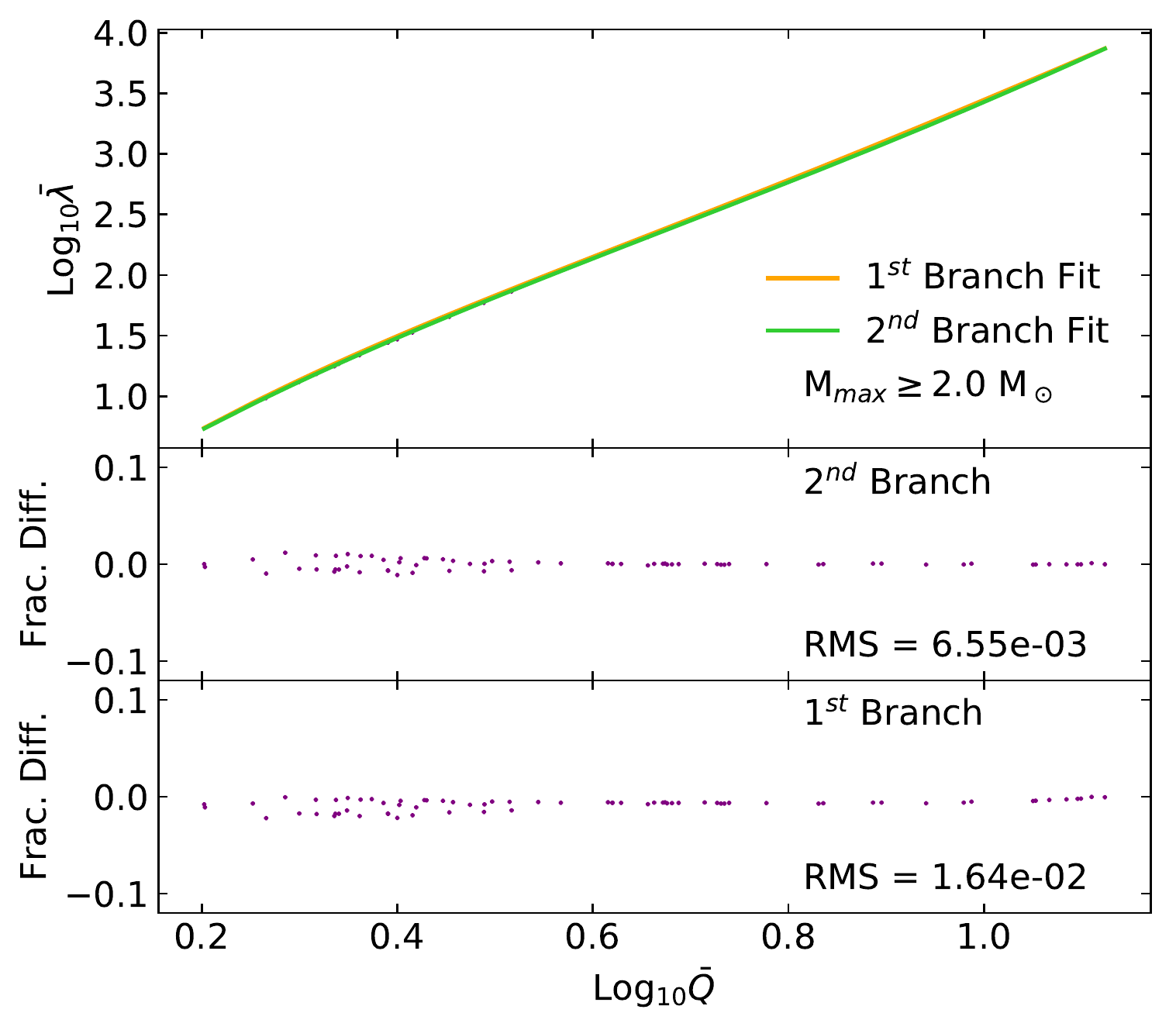}
    \caption{The $\log_{10}{\bar Q}-\log_{10}{\bar\lambda}$ relation and its residuals.  The panels correspond to the panels in Figure~\ref{fig:QIPlots}.  We find, in agreement with \cite{2013PhRvD..88b3009Y}, that the $\bar{Q}$-$\bar{\lambda}$ relation is not as tight as the $\bar{Q}$-$\bar{I}$ relation. This statement is also true when we focus exclusively on neutron stars in the second stable branch of central density.}
    \label{fig:QLPlots}
\end{figure*}

In Figures~\ref{fig:QIPlots} and \ref{fig:QLPlots} we present the fits and residuals for various parts of the data.  Figure~\ref{fig:QIPlots} shows $\log_{10}{\bar Q}$ versus $\log_{10}{\bar I}$ for our first-branch stars; for our first-branch stars constructed using EoS with a maximum mass greater than $2~M_\odot$; for all second-branch neutron stars; and for second-branch neutron stars constructed using EoS with a maximum mass greater than $2~M_\odot$.  We perform the mass cuts to gain insight into the universal relations; of course, EoS with maximum masses below $2~M_\odot$ are disfavored by the observations of high-mass neutron stars (see \cite{2010Natur.467.1081D,2013Sci...340..448A,2016ApJ...832..167F,2018ApJ...859...47A,2020NatAs...4...72C,2020arXiv200614601M}).  In the panels with second-branch neutron stars we show the residuals to a fit to just those stars, and the residuals to a fit constructed using only the first-branch stars.  We found that the greatest outliers tended to have a polytropic index $\sim 5$ (i.e., the maximum allowed) in the first density interval (from a baryonic rest mass density $\rho=\rho_s/2$ to $\rho=\rho_s$) and then a sharp drop of polytropic index to $\sim 2-3$ in the next two intervals.  

Figure~\ref{fig:QLPlots} follows the same pattern, but for $\log_{10}{\bar Q}$ versus $\log_{10}{\bar\lambda}$.  We find that both the Q-I and Q-$\lambda$ relations become tighter when the maximum mass cut is stricter.  This trend also applies to the second-branch neutron stars (see \cite{2018PhRvD..97h4038P}, but also see \cite{2018EPJA...54...26B}).  This trend with increasing maximum mass may be related to the observation that greater compactness tightens the relation \cite{2013Sci...341..365Y,2014PhRvD..90f3010Y}, but it is not identical: our central densities are picked randomly, which means that from an EoS with a high maximum mass we can select a star with low mass and low compactness.  We find that the second-branch relations are slightly different from the first-branch relations; this is evident from the panels showing the second-branch data, where the first-branch fits to the second-branch data have a larger root-mean-squared (rms) spread, and a larger maximum deviation, than the fits directly to the second-branch data.  Overall, and consistent with \cite{2013PhRvD..88b3009Y}, we find that the ${\bar Q}-{\bar\lambda}$ relation is not quite as tight as the ${\bar Q}-{\bar I}$ relation.

Table~\ref{tab:fits_nonA} gives our first-branch best-fit parameters, and Table~\ref{tab:fits_Anom} does the same for our second-branch best-fit parameters. 

\begin{table}[h]
\caption{\label{tab:fits_nonA}Fitting parameters of Equation~(\ref{eq:4orderfit}) for the relations between Q and I, $\lambda$, $M\omega_R$, and $M\omega_I$ for the first-branch stars. In the top four rows we require $M_{\rm max}>1.0~M_\odot$, and in the bottom four rows we require $M_{\rm max}>2~M_\odot$.}
\resizebox{.49\textwidth}{!}{
\begin{ruledtabular}
\begin{tabular}{cccccc}
1$^{\rm st}$ branch EoS Fits & $c_0$  & $c_1$    & $c_2$  & $c_3$   & $c_4$ \\ \hline
I-Q from \cite{2017PhR...681....1Y} & 0.6050 & 0.2376 & 0.0132 & 0.0084 & 0.0002\\
I-Q                & 0.5995 & 0.3000   & 0.1468 & 0.0198  & 0.0242\\ 
Love-Q             & -0.4969 & 4.3256  & -2.6411 & 1.7659  & -0.3613\\ 
$M\omega_R$-Q      & 0.4602 & 0.1632 & -0.7951 & 0.5854 & -0.1575\\ 
$M\omega_I$-Q      & -0.0266 & 0.8640 & -1.0778 & 0.6691 & -0.1941\vspace{6pt}
\\ 
I-Q & 0.5971 & 0.3158   & 0.1141 & 0.0514  & 0.0130\\ 
Love-Q & -0.5053 & 4.3761   & -2.7387 & 1.8528  & -0.3901\\ 
$M\omega_R$-Q & 0.4605 & 0.1593 & -0.7887 & 0.5839 & -0.1585\\ 
$M\omega_I$-Q & -0.0273 & 0.8730 & -1.1051 & 0.6955 & -0.2017\\ 
\end{tabular}
\end{ruledtabular}}
\end{table}

\begin{table}[h]
\caption{\label{tab:fits_Anom}Fitting parameters of Equation~(\ref{eq:4orderfit}) for the relations between Q and I, $\lambda$, $M\omega_R$, and $M\omega_I$ for second-branch stars. In the top four rows we require $M_{\rm max}>1.0~M_\odot$, and in the bottom four rows we require $M_{\rm max}>2~M_\odot$.}
\resizebox{.49\textwidth}{!}{
\begin{ruledtabular}
\begin{tabular}{cccccc}
2$^{\rm nd}$ branch EoS Fits & $c_0$  & $c_1$    & $c_2$  & $c_3$   & $c_4$ \\ \hline
I-Q                & 0.5714 & 0.4799   & -0.3404 & 0.4863  & -0.1222\\
Love-Q             & -0.5211 & 4.5361  & -3.4479 & 2.6458  & -0.6572\\
$M\omega_R$-Q      & 0.4244 & 0.4790 & -1.5159 & 1.2196 & -0.3536\\
$M\omega_I$-Q      & -0.0331 & 0.7823 & -0.6350 & 0.0673 & 0.0549 \vspace{6pt}\\ 
I-Q & 0.6223 & 0.1681   & 0.3143 & -0.0500  & 0.0283\\ 
Love-Q & -0.4728 & 4.1735   & -2.6090 & 1.9531  & -0.4698\\
$M\omega_R$-Q & 0.4170 & 0.5965 & -2.0125 & 1.9175 & -0.6632\\
$M\omega_I$-Q & -0.0523 & 0.9547 & -1.1088 & 0.5688 & -0.1265\\ 
\end{tabular}
\end{ruledtabular}}
\end{table}

\subsection{Correlations with $w$-mode frequencies}
\label{sec:w-mode}
          
        \begin{figure*}[ht]
                \includegraphics[width=0.49\textwidth]{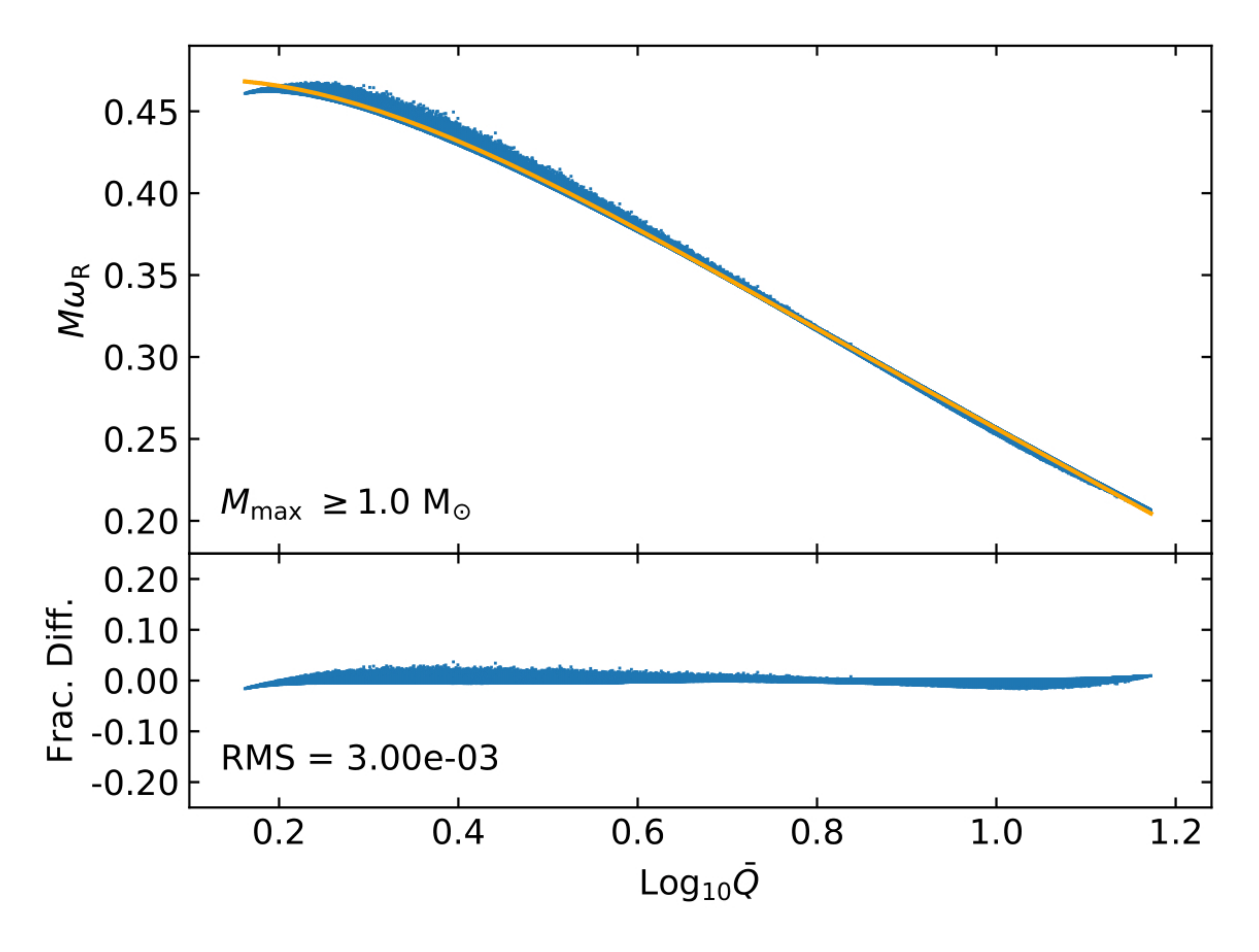}
                \includegraphics[width=0.49\textwidth]{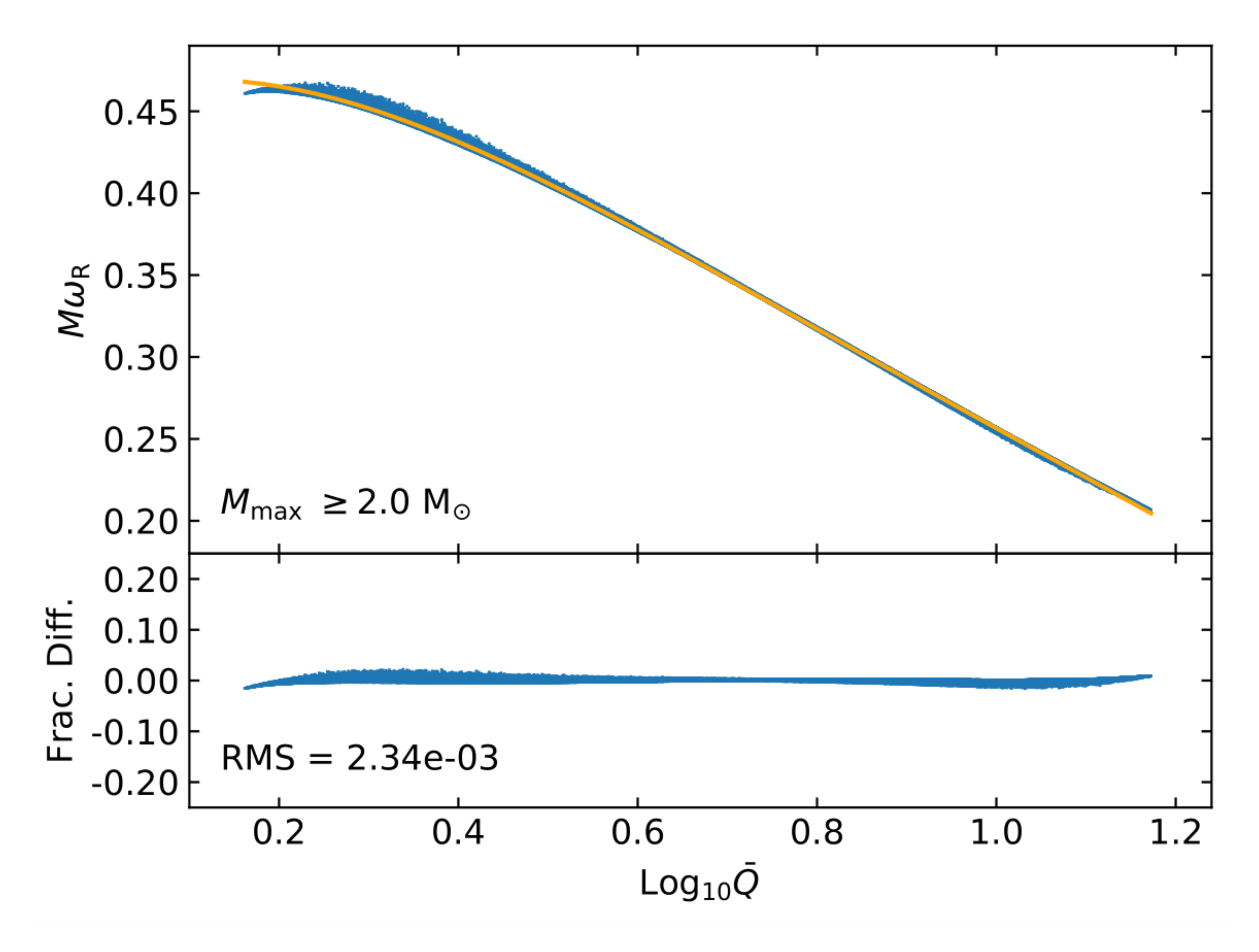}
                \includegraphics[width=0.49\textwidth]{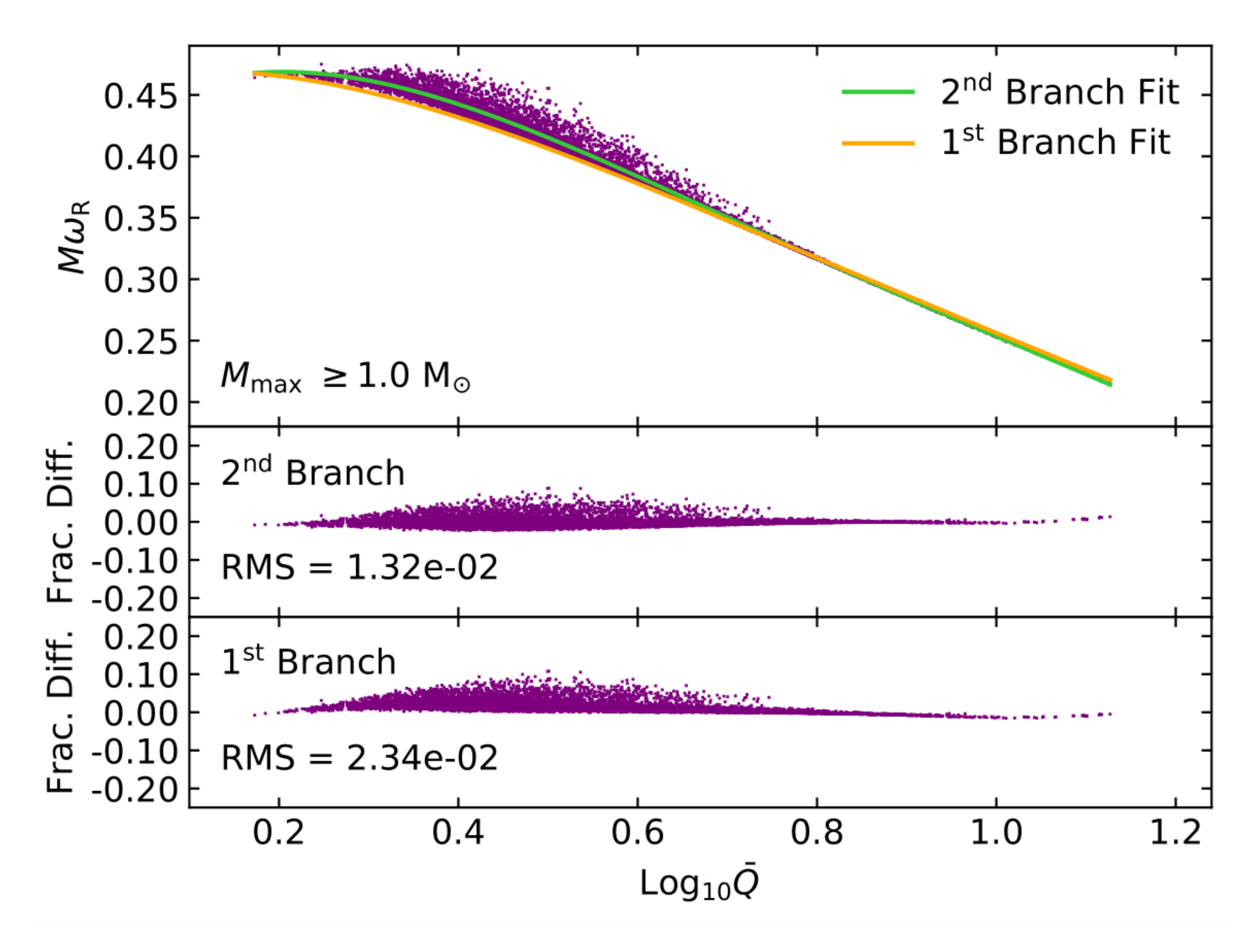}
                \includegraphics[width=0.49\textwidth]{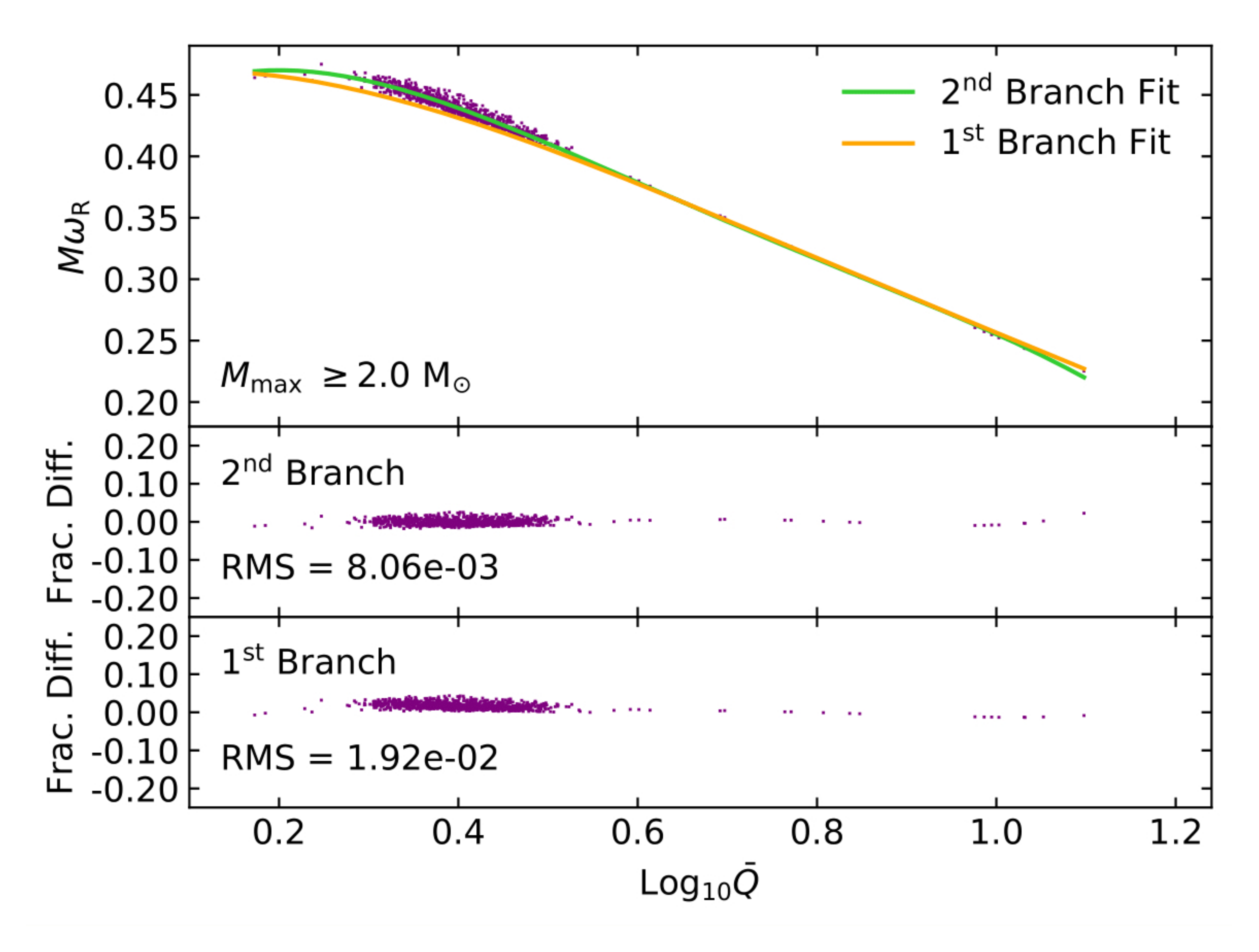}
            \caption{The relation between the real component $\omega_R$ of the frequency of the $w$-mode, and $\log_{10}{\bar Q}$.  The panels correspond to those of Figure~\ref{fig:QIPlots}.  The correlations are tightened when we impose a lower limit on the maximum mass, and the second-branch fit to the second-branch data is tighter than the first-branch fit to the second-branch data.  However, the tightening is not as pronounced as it is for the ${\bar Q}-{\bar\lambda}$ and ${\bar Q}-{\bar I}$ relations.}
            \label{fig:wr_vs_Q}
        \end{figure*}

        \begin{figure*}[ht]
                \includegraphics[width=0.49\textwidth]{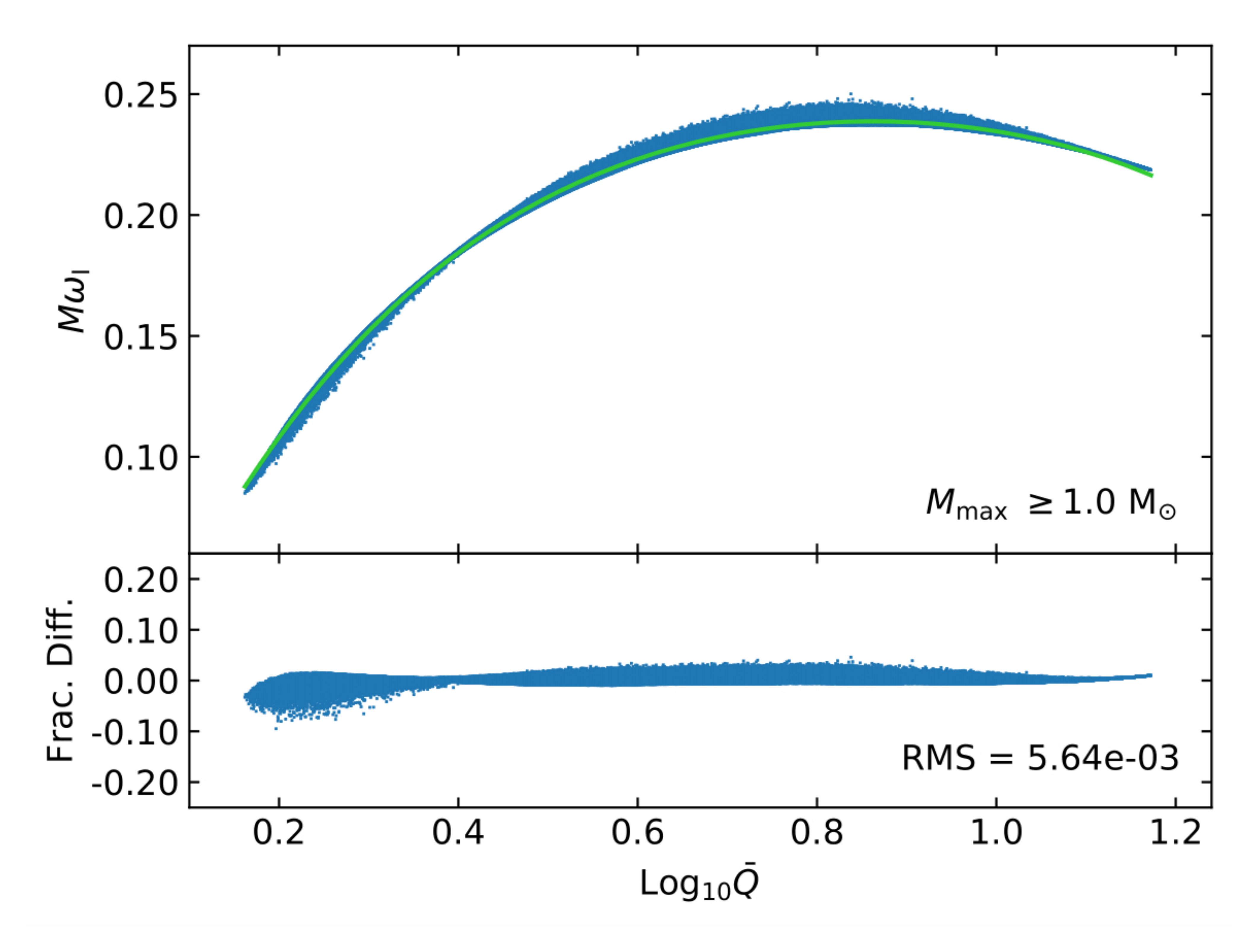}
                \includegraphics[width=0.49\textwidth]{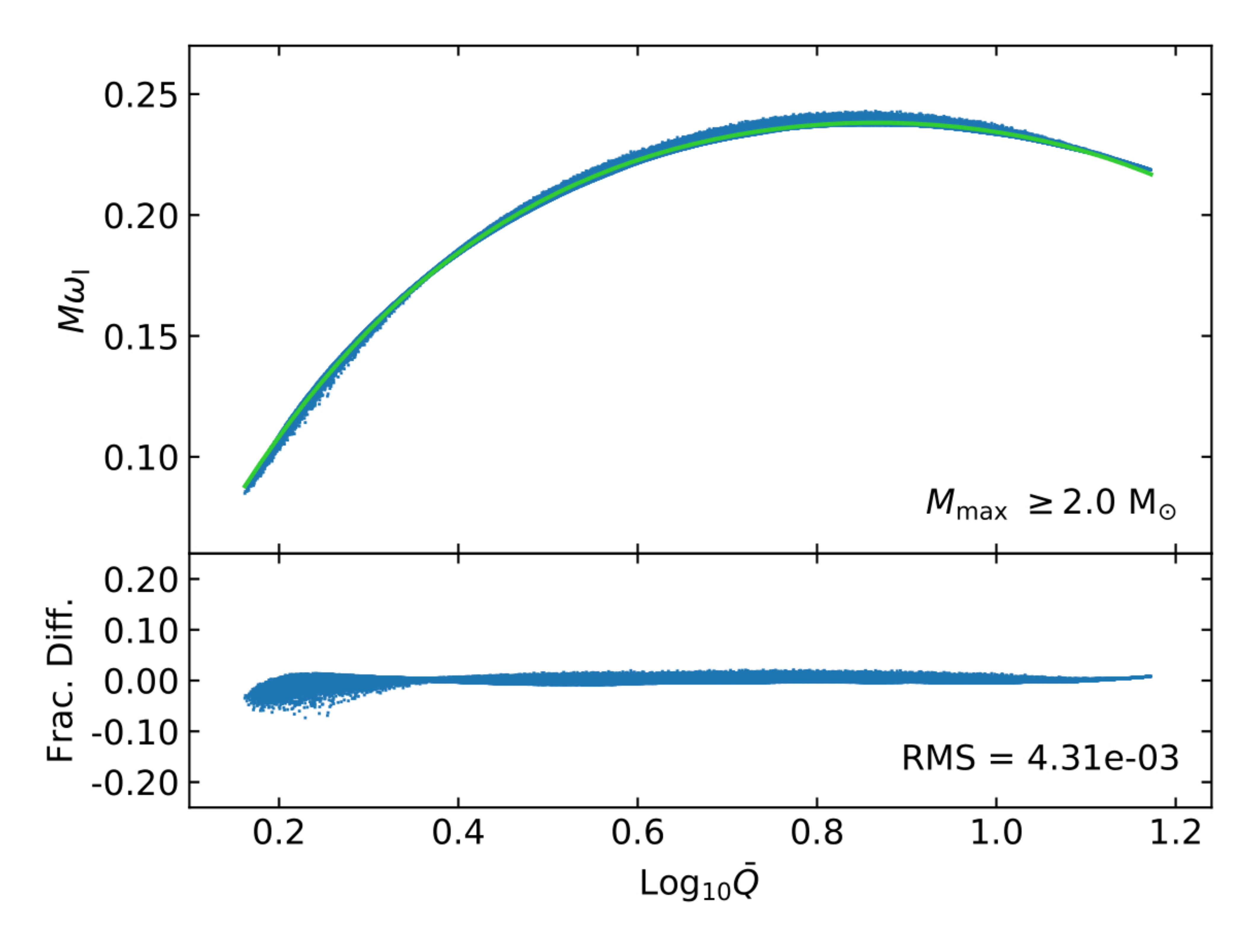}
                \includegraphics[width=0.49\textwidth]{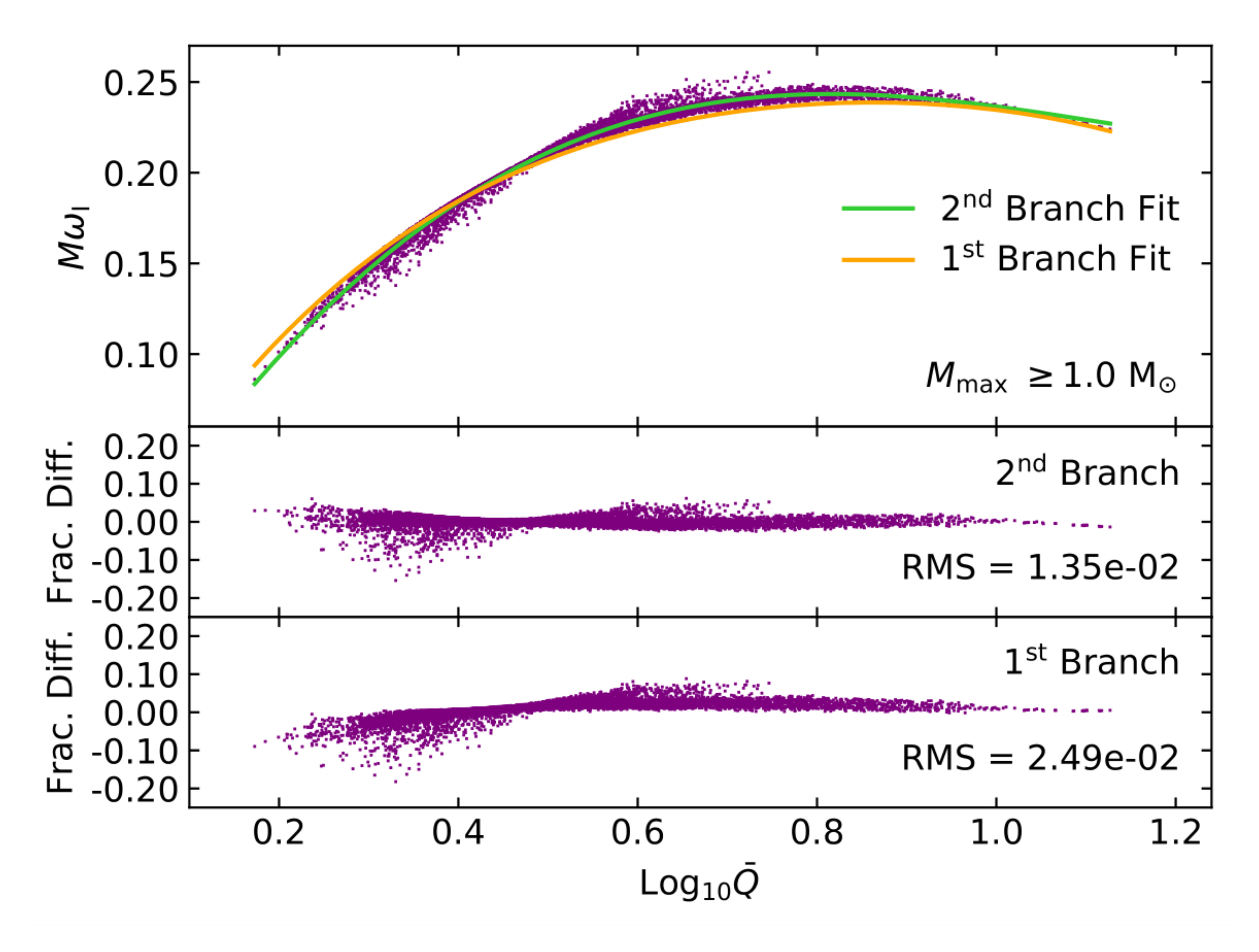}
                \includegraphics[width=0.49\textwidth]{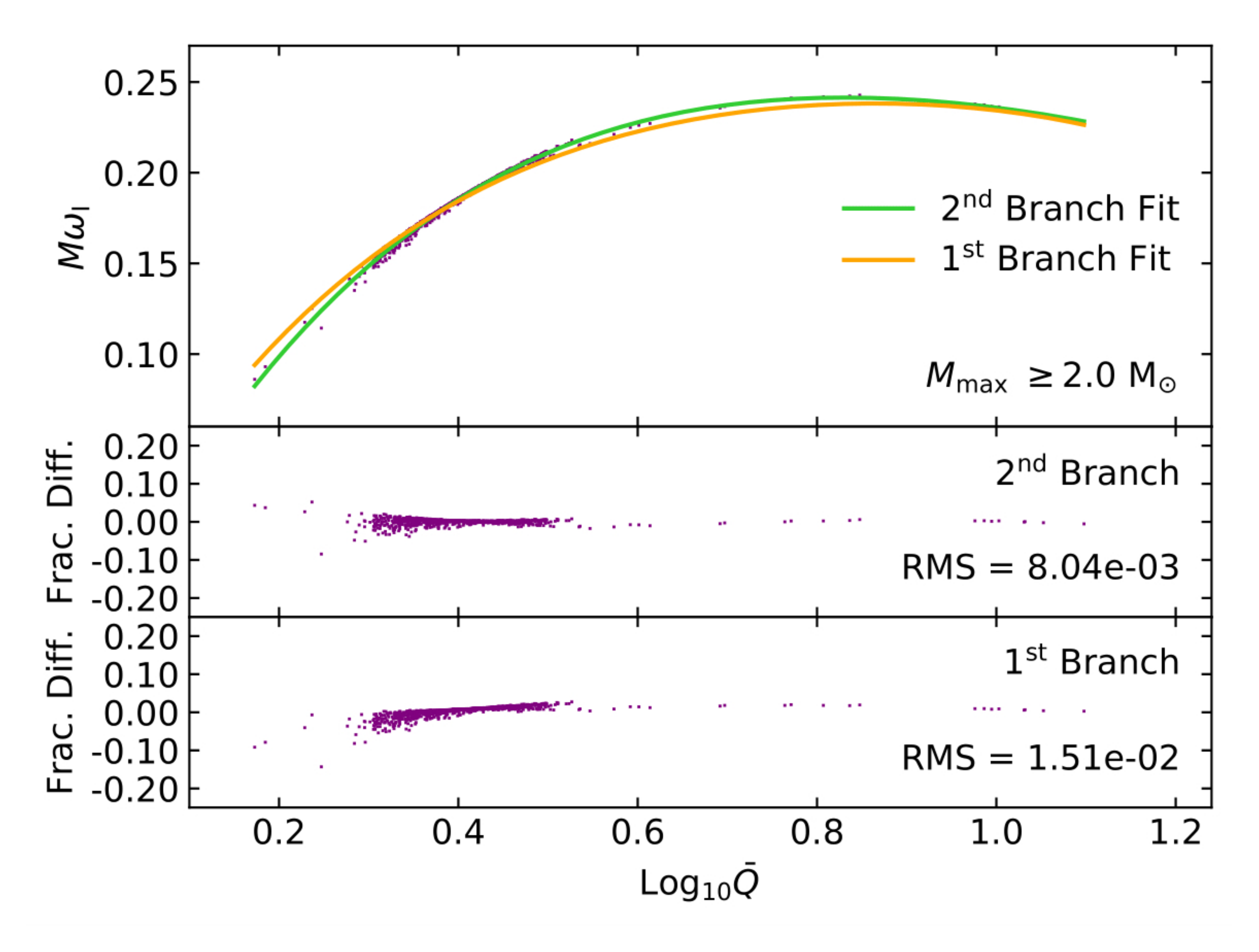}
            \caption{The relation between the imaginary component $\omega_I$ of the frequency of the $w$-mode, and $\log_{10}{\bar Q}$.  The panels correspond to those of Figure~\ref{fig:QIPlots}.  The residuals show significant systematics; for example, the spread at low ${\bar Q}$ is much greater than at higher ${\bar Q}$. }
            \label{fig:wi_vs_Q}
        \end{figure*}

Previous work has demonstrated that $f$-mode frequencies correlate well with the inverse square root of ${\bar I}$ \cite{2010ApJ...714.1234L,2015PhRvD..91d4034C}.  In this section we investigate similar relations involving the $w$-mode frequencies.

$w$-modes excite very little fluid motion; therefore they are expected to follow universal relations, even though earlier attempts at a universal description still showed significant EoS-dependence.  In Figures~\ref{fig:wr_vs_Q} and \ref{fig:wi_vs_Q}, we show the relations between ${\bar Q}$ and the real (Figure~\ref{fig:wr_vs_Q}) and imaginary (Figure~\ref{fig:wi_vs_Q}) parts of the fundamental axial $w$-mode frequency, scaled with the NS mass to make them dimensionless. The panels follow the same pattern as in Figure~\ref{fig:QIPlots} and Figure~\ref{fig:QLPlots}.  We see that these relations are only slightly less tight than the I-Love-Q relations, and are thus also nearly independent of the high-density EoS.  

As with I-Love-Q we find that the second-branch relations are not as tight as the first-branch relations, and that the second branch deviates somewhat from the first branch.  The clear systematics in the residuals presented in Figures \ref{fig:wr_vs_Q} and \ref{fig:wi_vs_Q} indicate that the actual functional form of the $w$-mode universal relations deviates from the fourth-order polynomial assumed here. An analytical treatment following the series expansion presented in \cite{2020PhRvD.101l4006J} might provide a better description. 
        
\section{Conclusions}
\label{sec:conclusions}

Our study of a large number of parameterized high-density EoS for neutron stars confirms the tightness of the I-Love-Q relation, and establishes that the $w$-mode frequencies are also closely linked with I, $\lambda$, and $Q$.  We show that for $I$, $\lambda$, $Q$, and $w$, second-branch stars follow a slightly different and somewhat less tight relation than the first-branch stars.

Currently none of $I$, $\lambda$, $Q$, or the $w$-mode complex frequency has been measured for any neutron star.  Perhaps the closest is $\lambda$, for which we have an interesting upper limit from GW170817 \cite{2017PhRvL.119p1101A,2018PhRvL.121i1102D,2018PhRvL.121p1101A} and a less constraining upper limit from GW190425 \cite{2020ApJ...892L...3A}, and which could be measured during a future, very high signal-to-noise, gravitational wave event.  It is also hoped that $I$ could be measured from a binary pulsar system \cite{2005ApJ...629..979L}, although initial optimism has given way to understanding that there are numerous practical difficulties with this measurement.  In principle, observations of a highly eccentric double neutron star coalescence using a third-generation gravitational detector could yield $I$ and $\lambda$ \cite{2017ApJ...837...67C}.

Because $w$-modes have frequencies $\sim 5-12$~kHz depending on the EoS \cite{1999LRR.....2....2K}, their detection likely will require specialized instruments attuned to such high frequencies.  One proposal for the next decade is the development of a ground-based detector with an enhanced high-frequency sensitivity in a range between 900~Hz and 5~kHz, through the project ``OzHF'' \citep{2020CQGra..37gLT02A}.  It has also been suggested that $w$-mode frequencies as low as $\sim 2-3$~kHz, along with $f$-mode frequencies, could be observed from protoneutron stars during the early stages of core-collapse supernovae \cite{2017PhRvD..96f3005S}.

In summary, determination of any of $I$, $\lambda$, $Q$, or the $w$-mode frequencies, let alone more than one of them, will be challenging.  Nonetheless, the robustness of the relations means that measurements have broad implications for the theory of strong gravity and for optimally precise inferences of properties of neutron stars.

\acknowledgements

We thank David Blaschke, Milva Orsaria, Iara Ota, Ignacio Ranea, Luciano Rezzolla, Kent Yagi, and Nico Yunes for helpful suggestions on an earlier version of our manuscript.  V. G. acknowledges support from the Brazilian National Council for Scientific and Technological Development (CNPq) and the S{\~a}o Paulo Research Foundation (FAPESP) through Grant 2019/20740-5.  M. C. M. thanks the Radboud Excellence Initiative for supporting his stay at Radboud University during part of this work.  C. C. acknowledges support from NASA under Grant No. 80GSFC17M0002. C. C. and M. C. M. are grateful for the hospitality of Perimeter Institute where part of this work was carried out. Research at Perimeter Institute is supported in part by the Government of Canada through the Department of Innovation, Science and Economic Development Canada, and by the Province of Ontario through the Ministry of Colleges and Universities. This research was also supported in part by the Simons Foundation through the Simons Foundation Emmy Noether Fellows Program at Perimeter Institute.

\end{document}